\documentclass[aps,superscriptaddress,eqsecnum,nofootinbib,preprintnumbers]{revtex4}
\usepackage{amsmath}
\usepackage{amsfonts}
\usepackage{amssymb}
\usepackage{latexsym}
\usepackage[dvips]{graphicx}
\usepackage{multirow}

\usepackage{graphicx,epsfig}
\usepackage{amsmath}
\usepackage{amssymb}
\usepackage{subfigure}

\usepackage{relsize}
\usepackage{textcomp}

\def\be{\begin{equation}}
\def\ee{\end{equation}}

\begin{document}

\title{IR Quantum Gravity solves naturally cosmic acceleration and its coincidence problem}
\thanks{Essay has been awarded an Honorable Mention in the Gravity Essay Competition 2019 with
the Essay published in a Special issue of IJMPD.}
\date{\today}

\author{Fotios K. Anagnostopoulos}
\email{fotis-anagnostopoulos@hotmail.com}
\affiliation{National and Kapodistrian University of Athens, Physics Department,
Panepistimioupoli Zografou, 15772, Athens, Greece}

\author{Georgios Kofinas}\thanks{Corresponding author}
\email{gkofinas@aegean.gr}
\affiliation{Research Group of Geometry, Dynamical Systems and Cosmology,\\
Department of Information and Communication Systems Engineering,\\
University of the Aegean, Karlovassi 83200, Samos, Greece}

\author{Vasilios Zarikas}
\email{vzarikas@uth.gr}
\affiliation{School of Engineering, Nazarbayev University, Nur-Sultan (Astana), Kazakhstan}
\affiliation{General Department, University of Thessaly, Lamia, Greece}

\begin{abstract}

The novel idea is that the undergoing accelerated expansion of the universe happens due to
infrared quantum gravity modifications at intermediate astrophysical scales of galaxies or
galaxy clusters, within the framework of Asymptotically Safe gravity. The reason is that
structures of matter are associated with a scale-dependent positive cosmological constant of
quantum origin. In this context no extra unproven energy scales or fine-tuning are used.
Furthermore, this model was confronted with the most recent observational data from a variety of
probes, and with aid of Bayesian analysis, the most probable values of the free parameters were
extracted. Finally, the model proved to be statistically equivalent with $\Lambda$CDM, and thus being
able to resolve naturally the concept of dark energy and its associated cosmic coincidence problem.

\end{abstract}

\maketitle

\section{Introduction}
\label{Introduction}

The idea that Quantum Gravity effects can be important at astrophysical and cosmological distances
has recently attracted much attention. In particular the framework of Exact Renormalization Group
(RG) approach for quantum gravity \cite{Reuter:1996cp}, also found in the literature under the
names Asymptotic Safety (AS) or Quantum Einstein Gravity, has opened the possibility of
investigating both the ultraviolet (UV) and the infrared (IR) sector of gravity in a systematic
manner.

The key element is the Effective Average Action $\Gamma_k[g_{\mu\nu}]$, a Wilsonian coarse-grained
free energy, which defines an effective field theory appropriate for the momentum scale $k$.
This $\Gamma_{k}$, evaluated at tree level, describes appropriately all gravitational phenomena,
including all loop effects. The application to Einstein-Hilbert action generates RG flow
equations \cite{Dou:1997fg}, which have made possible the consistent study of the scaling behavior
of Newton constant $G$ and cosmological constant $\Lambda$ at high energies \cite{Litim:2003vp},
\cite{Bonanno:2004sy}. The initial idea was first demonstrated by Weinberg \cite{wein2}, where
he suggested that a pertubatively divergent theory could be consistently defined in four dimensions
at a nontrivial UV fixed point with the dimensionless $g(k)=G(k) k^2$ non-vanishing in the
$k\rightarrow\infty$ limit. In the framework of AS, one should also include $\Lambda$, which
becomes energy dependent and receives quantum contribution form vacuum fluctuations.

Recent works have also considered matter fields or a growing number of purely
gravitational operators in the action. In particular, truncations involving quadratic terms in
the curvature or higher powers of the Ricci scalar have been studied \cite{Benedetti:2009gn},
\cite{Codello:2008vh}. In all the investigations the UV critical surface has turned out to be
finite dimensional ($d_{\rm UV}=3$), implying that the theory is nonperturbatively renormalizable.

A weakly coupled gravity at high energies is expected to generate important consequences in
several astrophysical and cosmological contexts and in fact the RG flow of $\Gamma_k$, obtained
by different truncations of theory space, has been the basis of various investigations of
``RG improved'' black hole spacetimes \cite{Bonanno:2006eu}, \cite{Kofinas:2015sna}
and early Universe models \cite{2006rdgp.conf.461B}-\cite{Zarikas:2018wfv}.

\section{IR point from AS approach}

The behavior of AS theory is more complicated at low energies, corresponding at cosmological or
astrophysical scales. The problem arises because the $\beta$-functions of any local operator of
the type $\sqrt{g}R^n$ are singular in the IR due to the presence of a pole at
$\lambda(k)=\Lambda(k)/k^2=1/2$. This pole indicates that the Einstein-Hilbert truncation is not
trustable approximation and new relevant operators emerge in the $k\rightarrow 0$ limit. It was
shown in \cite{Reuter:2004nx} that the dynamical origin of these strong IR effects is due to the
``instability driven renormalization'', a phenomenon well-known from other physical systems
\cite{Alexandre:1998ts}-\cite{Bonanno:2004pq}. Fortunately, the low energy domain of the theory
is regulated by an IR fixed point which drives the cosmological constant to zero. Out of that,
some first astrophysical consequences and a possible explanation for the galaxy rotation curves
without dark matter appeared \cite{Reuter:2004nx}, \cite{Reuter:2003ca}, \cite{Reuter:2004nv},
although a detailed analysis based on available experimental data is still missing.

Very recently, authors in \cite{Kofinas:2017gfv}, \cite{Anagnostopoulos:2018jdq} showed that the
coincidence problem of cosmic acceleration can be explained naturally and without introducing new
energy scales or fine-tuning in the context of a Swiss cheese cosmological model which utilizes
this IR behavior of AS gravity. The present essay focuses on this discovery. The importance of these
works lies on the fact that no special ``dark energy'' field is required and the explanation uses
only the concrete well motivated framework of AS.

More precisely, in the proposed scenario, including all the components of the metric fluctuation
$h_{\mu\nu}$ \cite{Bonanno:2004sy}, quantum effects dynamically drive $\Lambda$ along the RG
trajectories generated by the unstable infrared modes of the gravitational sector. Close to the
IR fixed point, $\Lambda$ runs proportional to $k^{2}$ to avoid the singularity,
\begin{equation}
\frac{\Lambda(k)}{k^2}=\lambda^{\mathrm{IR}}_{*}+h_2 k^{2\theta}\,\,, \;\;\;\;
k\rightarrow 0\,,
\label{ir}
\end{equation}
where $\lambda^{\mathrm{IR}}_{*} < 1/2$ is an infrared fixed point of the $\lambda$-evolution,
$h_2$ is a constant related to the eigenvalues of the stability matrix, and $\theta>0$ is a
critical exponent which parametrizes the subleading terms and ensures that the fixed point is
attractive. According to (\ref{ir}), the renormalized $\Lambda$ vanishes at
very large (cosmological) distances, $\Lambda (k \to 0)=0$, regardless of its bare value.
Furthermore, recent investigations based on a conformal reduction of Einstein gravity discovered
a new IR fixed point suggesting the existence of the counterpart of the physical IR fixed point
present in the full theory \cite{Manrique:2010am}.

\section{Cosmology by matching local astrophysical metrics with global cosmological metric}

In \cite{Kofinas:2017gfv}, the recent cosmic acceleration naturally emanated from the recent
formation of structure. A Swiss cheese (Einstein-Strauss) model was constructed to derive the
cosmology. The interior static spherically symmetric metric, modeling a galaxy or a galaxy cluster,
matches smoothly to a cosmological exterior across a spherical boundary. A quantum
improved Schwarzschild-de Sitter (SdS) interior metric was used, which contains the appropriate
antigravity effect. This model uses dimensionless order one parameters of AS, the conventional
Newton constant $G_{\!N}$ and the astrophysical length scale. It provides a recent passage to
the sufficient acceleration, while the freedom of the order one parameters has to be
constrained by observational data. To the best of our knowledge, this is the first solution of the
dark energy problem without using fine-tuning or introducing add-hoc energy scales.

The exterior FRW metric is
\begin{equation}
ds^2=-dt^2+a^2(t)\left[\frac{dr^2}{1-\kappa r^2}+r^2\left(d\theta^2+\sin^2\!\theta \,d\varphi^2\right)
\right]
\label{eq:FRW}
\end{equation}
with $a(t)$ the scale factor. The interior metric has the form
\begin{equation}
ds^2=-\Big(1-\frac{2G_k M}{R}-\frac{1}{3}\Lambda_k R^2\Big)dT^2
+\frac{dR^2}{1-\frac{2G_k M}{R}-\frac{1}{3}\Lambda_k R^2}
+R^2\left(d\theta^2+\sin^2\theta d\varphi^2\right)
\label{ASBH}
\end{equation}
with the functions $G_{k}=G(k), \Lambda_{k}=\Lambda(k)$ determined by AS. The matching of the
two patches occurs at the boundary with constant $r=r_{\Sigma}$, which at the same time
experiences the universal expansion. Since astrophysical structures are still large compared
to the cosmological scales, $\Lambda(k)$ is expected to differ slightly from its IR form $k^{2}$,
so a power law $\Lambda_{k}=\gamma k^{b}$, with $\gamma>0$ and $b$ close to the value 2, is a fair
approximation of the running behavior (\ref{ir}). Additionally, it is set $G_{k}=G_{N}$ at
observable macroscopic distances, in agreement with standard Newton law.

The energy scale $k$ is expected to be associated with a characteristic length scale $L$, $k=\xi/L$,
where $\xi$ is an order one dimensionless number. In the Swiss cheese approach, only the value
$R_{S}=ar_{\Sigma}$ of the Schucking radius (or $k_{S}$) enters the cosmic evolution. Although
a simple option is $L=R$, a more natural one is to set as $L$ the proper distance
\begin{equation}
D(R)=\int^{R}\!d\mathcal{R}\,\Big(1-\frac{2G_N M}{\mathcal{R}}-
\frac{1}{3}\Lambda_k \mathcal{R}^2\Big)^{-1/2}\,.
\label{ekryw}
\end{equation}

The new cosmological constant term is
$\frac{1}{3}\Lambda_{k}R^{2}\sim \frac{1}{G_{\!N}}
\big(\frac{\sqrt{G_{\!N}}}{r_{\Sigma}}\big)^{\!b}\,\big(\frac{r_{\Sigma}}{D}\big)^{\!b}R^{2}$,
where $\tilde{\gamma}=\gamma G_{\!N}^{1-\frac{b}{2}}$ a dimensionless order one number.
For $b$ close to the value 2.1, the quantity
$\frac{1}{G_{\!N}}\big(\frac{\sqrt{G_{\!N}}}{r_{\Sigma}}\big)^{\!b}$ is very close to the order
of magnitude of the standard cosmological constant $\Lambda\simeq 4.7\times10^{-84}\text{GeV}^{2}$ of
the concordance $\Lambda$CDM model, while the factor $\big(\frac{r_{\Sigma}}{D}\big)^{\!b}$
contributes only a small distance-dependent deformation since the today value of $D_{S}$ should
be of order $r_{\Sigma}$ in order to have the correct amount of dark energy. Therefore, the hard
coincidence of the standard $\Lambda\sim H_{0}^{2}$, has been exchanged with a mild
adjustment of the index $b$ close to 2.1.

The Hubble evolution in terms of the redshift $z$ arises by the integration of the cosmological
equations of the model which consider the Israel-Darmois matching conditions on the boundary radius,
\begin{equation}
\frac{H^{2}(z)}{H_{0}^{2}}=\Omega_{m0}(1\!+\!z)^{3}
+\Big[\Omega_{DE0}^{-\frac{1}{b}}-\frac{3^{\frac{1}{b}}}{\xi\tilde{\gamma}^{\frac{1}{b}}}
(G_{\!N}H_{0}^{2})^{\frac{1}{b}}\frac{r_{\Sigma}a_{0}}{\sqrt{G_{\!N}}}\,\frac{z}{1\!+\!z}\Big]^{-b}
+\Omega_{\kappa 0}(1\!+\!z)^{2}\,.
\label{kieb}
\end{equation}

In addition, a preliminary analysis has shown that there is no obvious
contradiction between the model discussed and the internal dynamics of the astrophysical object.
The potential and the force due to the varying cosmological constant term are small percentages
of the corresponding Newtonian potential and force, where the precise values depend on the
considered structure and the considered point at the boundary of the object or inside.

\section{Thorough test of the model using observational data}

No matter what are the theoretical merits of a given model, it needs to be confronted with the
observational data. In \cite{Anagnostopoulos:2018jdq}, authors used the most recent observational
data sets, namely direct measurements of the Hubble rate $H(z)$ \cite{Jimenez:2001gg}, Supernovae
Ia (Pantheon data set \cite{Scolnic:2017caz}), Quasi-Stellar-Objects (QSO), Baryonic Acoustic
Oscillations and direct measurements of the CMB shift parameters, to constraint the free parameters
of the AS cosmological model \cite{Kofinas:2017gfv}. It was found that this model is
very efficient and in excellent agreement with observations. The energy density of matter that was
calculated is compatible with the same quantity imposed by concordance cosmology from the CMB
angular power spectrum. Another result is that the AS model supports a lower value of Hubble
constant than the value derived from Cepheids, so the best fit value for $H_0$ is closer to the
Planck value.

In addition, after reconstructing the effective dark energy equation-of-state parameter $w_{DE}(z)$
using the derived values of the free parameters, it was found that the today's value $w_{DE0}$ is
close to $w=-1$. Both the transition redshift and the current value of the deceleration parameter
are in perfect agreement with the corresponding values calculated with model-independent techniques.

\begin{figure}
\includegraphics[width=0.5\textwidth]{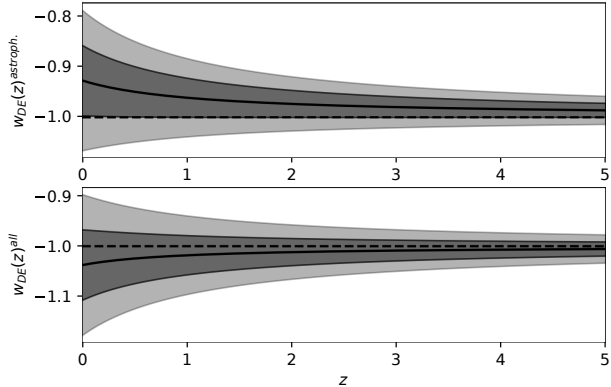}
\caption{The evolution of $w_{DE}(z)$ with the corresponding $1\sigma$ and
$2\sigma$ uncertainties. In the upper panel we use the best fit values
from the combination $H(z)$/Pantheon/QSO, while in the lower panel we
utilize those of $H(z)$/Pantheon/QSO/CMB$_{\rm shift}$.
The dashed line corresponds to $\Lambda$CDM value $w=-1$.}
\label{fig:wz_reconstruction}
\end{figure}

\begin{table*}[!]
\tabcolsep 5.5pt
\vspace{5mm}
\begin{tabular}{ccccccccc} \hline \hline
Model & $\Omega_{m0}$ & $h$ & $\Omega_{b0}h^2$ &
$\chi_{\text{min}}^{2}$ & ${\rm AIC}$& $\Delta$AIC$ $\vspace{0.05cm}\\ \hline
%----------------------------------------
\hline
\\
\multicolumn{7}{c}{\emph{$H(z)$/\text{Pantheon}/\text{QSO}}}\\ \\
AS model & $0.270_{-0.018}^{+0.019}$ & $0.684_{-0.012}^{+0.013} $ & - & $ 84.463 $ &
92.889 & 0.937 \vspace{0.01cm}\\

$\Lambda$CDM & $0.281^{+0.016}_{-0.015}$ & $0.686 \pm 0.013$ & -  & 85.700 & 91.952 &
0 \vspace{0.45cm}\\ %%%-----------------------------

\multicolumn{7}{c}{\emph{$H(z)$/\text{Pantheon}/\text{QSO}/\text{CMB} \text{shift}}}\\ \\
%----------------------------------------
AS model & $0.303 \pm 0.001 $ & $ 0.685 \pm 0.009$ & $0.0223
\pm 0.0002$ & 89.774 & 100.419 & 1.653 \vspace{0.01cm}\\ %%

% % % % alls good up here!!
$\Lambda$CDM& $0.307^{+0.008}_{-0.007} $ & $0.679 \pm 0.006 $& $0.0223 \pm 0.0001$ & 90.340 &
98.766  & 0\vspace{0.45cm}\\ %%
%------------------------------------%%OK!!!

\hline\hline
\end{tabular}
\caption[]{The observational constraints on the density parameters $\Omega_{m0}$, $\Omega_{b0}$
and the statistical criteria values $\chi^{2}_{\rm min}$, AIC, $\Delta {\rm AIC}$ for the AS
cosmology and $\Lambda$CDM.
\label{tab:Results1}}
\end{table*}

Finally, the comparison of $\Lambda$CDM model with the considered AS model in
terms of the fitting properties, using a variety of information criteria, revealed that the AS
model is statistically equivalent with that of $\Lambda$CDM. This is a significant conclusion
since the AS model, unlike the majority of the cosmological models, does not include new fields
in nature or has fine-tuning problems. Therefore, it must be considered as a viable and efficient
alternative cosmological scenario towards explaining the recent accelerated expansion of the universe.
Interestingly, the two models are expected to have differences at the perturbation level.

\begin{acknowledgments}

G. Kofinas and V. Zarikas acknowledge the support of Orau Grant Number 110119FD4534,
``Quantum gravity at astrophysical scales''.

\end{acknowledgments}

%\bibliography{t2}

\begin{thebibliography}{10}

%\cite{Reuter:1996cp}
\bibitem{Reuter:1996cp}
  M.~Reuter,
  ``Nonperturbative evolution equation for quantum gravity'',
  Phys.\ Rev.\ D {\bf 57}, 971 (1998),
  doi:10.1103/PhysRevD.57.971
  [hep-th/9605030].
  %%CITATION = doi:10.1103/PhysRevD.57.971;%%
  %675 citations counted in INSPIRE as of 30 Mar 2019

%\cite{Dou:1997fg}
\bibitem{Dou:1997fg}
  D.~Dou and R.~Percacci,
  ``The running gravitational couplings'',
  Class.\ Quant.\ Grav.\  {\bf 15}, 3449 (1998),
  doi:10.1088/0264-9381/15/11/011
  [hep-th/9707239].
  %%CITATION = doi:10.1088/0264-9381/15/11/011;%%
  %225 citations counted in INSPIRE as of 30 Mar 2019

%\cite{Litim:2003vp}
\bibitem{Litim:2003vp}
  D.~F.~Litim,
  ``Fixed points of quantum gravity'',
  Phys.\ Rev.\ Lett.\  {\bf 92}, 201301 (2004),
  doi:10.1103/PhysRevLett.92.201301
  [hep-th/0312114].
  %%CITATION = doi:10.1103/PhysRevLett.92.201301;%%
  %380 citations counted in INSPIRE as of 30 Mar 2019

%\cite{Bonanno:2004sy}
\bibitem{Bonanno:2004sy}
  A.~Bonanno and M.~Reuter,
  ``Proper time flow equation for gravity'',
  JHEP {\bf 0502}, 035 (2005),
  doi:10.1088/1126-6708/2005/02/035
  [hep-th/0410191].
  %%CITATION = doi:10.1088/1126-6708/2005/02/035;%%
  %107 citations counted in INSPIRE as of 30 Mar 2019

\bibitem{wein2}
  S.~Weinberg,
  ``Ultraviolet divergences in quantum theories of gravitation'',
  in S.W. Hawking and W.~Israel, editors, {\em General Relativity: an
  Einstein Centenary Survey}, Cambridge University Press, 1979.

%\cite{Benedetti:2009gn}
\bibitem{Benedetti:2009gn}
  D.~Benedetti, P.~F.~Machado and F.~Saueressig,
  ``Taming perturbative divergences in asymptotically safe gravity'',
  Nucl.\ Phys.\ B {\bf 824}, 168 (2010),
  doi:10.1016/j.nuclphysb.2009.08.023
  [arXiv:0902.4630 [hep-th]].
  %%CITATION = doi:10.1016/j.nuclphysb.2009.08.023;%%
  %141 citations counted in INSPIRE as of 30 Mar 2019

%\cite{Codello:2008vh}
\bibitem{Codello:2008vh}
  A.~Codello, R.~Percacci and C.~Rahmede,
  ``Investigating the Ultraviolet Properties of Gravity with a Wilsonian Renormalization Group Equation'',
  Annals Phys.\  {\bf 324}, 414 (2009),
  doi:10.1016/j.aop.2008.08.008
  [arXiv:0805.2909 [hep-th]].
  %%CITATION = doi:10.1016/j.aop.2008.08.008;%%
  %373 citations counted in INSPIRE as of 30 Mar 2019

%\cite{Bonanno:2006eu}
\bibitem{Bonanno:2006eu}
  A.~Bonanno and M.~Reuter,
  ``Spacetime structure of an evaporating black hole in quantum gravity'',
  Phys.\ Rev.\ D {\bf 73}, 083005 (2006),
  doi:10.1103/PhysRevD.73.083005
  [hep-th/0602159].
  %%CITATION = doi:10.1103/PhysRevD.73.083005;%%
  %152 citations counted in INSPIRE as of 30 Mar 2019

%\cite{Kofinas:2015sna}
\bibitem{Kofinas:2015sna}
  G.~Kofinas and V.~Zarikas,
  ``Avoidance of singularities in asymptotically safe Quantum Einstein Gravity'',
  JCAP {\bf 1510}, no. 10, 069 (2015),
  doi:10.1088/1475-7516/2015/10/069
  [arXiv:1506.02965 [hep-th]].
  %%CITATION = doi:10.1088/1475-7516/2015/10/069;%%
  %8 citations counted in INSPIRE as of 30 Mar 2019

\bibitem{2006rdgp.conf.461B}
  A.~{Bonanno} and M.~{Reuter},
  ``A cosmology of the Planck era from the renormalization group for quantum gravity'',
  in I.~{Ciufolini}, E.~{Coccia}, M.~{Colpi}, V.~{Gorini}, and R.~{Peron}, editors,
  {\em Recent Developments in Gravitational Physics}, p. 461, 2006.

%\cite{Kofinas:2016lcz}
\bibitem{Kofinas:2016lcz}
  G.~Kofinas and V.~Zarikas,
  ``Asymptotically Safe gravity and non-singular inflationary Big Bang with vacuum birth'',
  Phys.\ Rev.\ D {\bf 94}, no. 10, 103514 (2016),
  doi:10.1103/PhysRevD.94.103514
  [arXiv:1605.02241 [gr-qc]].
  %%CITATION = doi:10.1103/PhysRevD.94.103514;%%
  %11 citations counted in INSPIRE as of 30 Mar 2019

%\cite{Zarikas:2018wfv}
\bibitem{Zarikas:2018wfv}
  V.~Zarikas and G.~Kofinas,
  ``Singularities and Phenomenological aspects of Asymptotic Safe Gravity'',
  J.\ Phys.\ Conf.\ Ser.\  {\bf 1051}, no. 1, 012028 (2018),
  doi:10.1088/1742-6596/1051/1/012028
  %%CITATION = doi:10.1088/1742-6596/1051/1/012028;%%

%\cite{Reuter:2004nx}
\bibitem{Reuter:2004nx}
  M.~Reuter and H.~Weyer,
  ``Quantum gravity at astrophysical distances?'',
  JCAP {\bf 0412}, 001 (2004),
  doi:10.1088/1475-7516/2004/12/001
  [hep-th/0410119].
  %%CITATION = doi:10.1088/1475-7516/2004/12/001;%%
  %152 citations counted in INSPIRE as of 30 Mar 2019

%\cite{Alexandre:1998ts}
\bibitem{Alexandre:1998ts}
  J.~Alexandre, V.~Branchina and J.~Polonyi,
  ``Instability induced renormalization'',
  Phys.\ Lett.\ B {\bf 445}, 351 (1999),
  doi:10.1016/S0370-2693(98)01491-9
  [cond-mat/9803007].
  %%CITATION = doi:10.1016/S0370-2693(98)01491-9;%%
  %79 citations counted in INSPIRE as of 30 Mar 2019

%\cite{Lauscher:2000ux}
\bibitem{Lauscher:2000ux}
  O.~Lauscher, M.~Reuter and C.~Wetterich,
  ``Rotation symmetry breaking condensate in a scalar theory'',
  Phys.\ Rev.\ D {\bf 62}, 125021 (2000),
  doi:10.1103/PhysRevD.62.125021
  [hep-th/0006099].
  %%CITATION = doi:10.1103/PhysRevD.62.125021;%%
  %30 citations counted in INSPIRE as of 30 Mar 2019

%\cite{Bonanno:2004pq}
\bibitem{Bonanno:2004pq}
  A.~Bonanno and G.~Lacagnina,
  ``Spontaneous symmetry breaking and proper time flow equations'',
  Nucl.\ Phys.\ B {\bf 693}, 36 (2004),
  doi:10.1016/j.nuclphysb.2004.06.003
  [hep-th/0403176].
  %%CITATION = doi:10.1016/j.nuclphysb.2004.06.003;%%
  %20 citations counted in INSPIRE as of 30 Mar 2019

%\cite{Reuter:2003ca}
\bibitem{Reuter:2003ca}
  M.~Reuter and H.~Weyer,
  ``Renormalization group improved gravitational actions: A Brans-Dicke approach'',
  Phys.\ Rev.\ D {\bf 69}, 104022 (2004),
  doi:10.1103/PhysRevD.69.104022
  [hep-th/0311196].
  %%CITATION = doi:10.1103/PhysRevD.69.104022;%%
  %120 citations counted in INSPIRE as of 30 Mar 2019

%\cite{Reuter:2004nv}
\bibitem{Reuter:2004nv}
  M.~Reuter and H.~Weyer,
  ``Running Newton constant, improved gravitational actions, and galaxy rotation curves'',
  Phys.\ Rev.\ D {\bf 70}, 124028 (2004),
  doi:10.1103/PhysRevD.70.124028
  [hep-th/0410117].
  %%CITATION = doi:10.1103/PhysRevD.70.124028;%%
  %113 citations counted in INSPIRE as of 30 Mar 2019

%\cite{Kofinas:2017gfv}
\bibitem{Kofinas:2017gfv}
  G.~Kofinas and V.~Zarikas,
  ``Solution of the dark energy and its coincidence problem based on local antigravity sources without fine-tuning or new scales'',
  Phys.\ Rev.\ D {\bf 97}, no. 12, 123542 (2018),
  doi:10.1103/PhysRevD.97.123542
  [arXiv:1706.08779 [gr-qc]].
  %%CITATION = doi:10.1103/PhysRevD.97.123542;%%
  %2 citations counted in INSPIRE as of 30 Mar 2019

%\cite{Anagnostopoulos:2018jdq}
\bibitem{Anagnostopoulos:2018jdq}
  F.~K.~Anagnostopoulos, S.~Basilakos, G.~Kofinas and V.~Zarikas,
  ``Constraining the Asymptotically Safe Cosmology: cosmic acceleration without dark energy'',
  JCAP {\bf 1902}, 053 (2019),
  doi:10.1088/1475-7516/2019/02/053
  [arXiv:1806.10580 [astro-ph.CO]].
  %%CITATION = doi:10.1088/1475-7516/2019/02/053;%%

%\cite{Manrique:2010am}
\bibitem{Manrique:2010am}
  E.~Manrique, M.~Reuter and F.~Saueressig,
  ``Bimetric Renormalization Group Flows in Quantum Einstein Gravity'',
  Annals Phys.\  {\bf 326}, 463 (2011),
  doi:10.1016/j.aop.2010.11.006
  [arXiv:1006.0099 [hep-th]].
  %%CITATION = doi:10.1016/j.aop.2010.11.006;%%
  %124 citations counted in INSPIRE as of 30 Mar 2019

%\cite{Jimenez:2001gg}
\bibitem{Jimenez:2001gg}
R.~Jimenez and A.~Loeb,
``Constraining cosmological parameters based on relative galaxy ages'',
Astrophys.\ J.\  {\bf 573}, 37 (2002),
doi:10.1086/340549
[astro-ph/0106145].
%%CITATION = doi:10.1086/340549;%%
%261 citations counted in INSPIRE as of 01 Mar 2018

\bibitem{Scolnic:2017caz}
D.~M.~Scolnic {\it et al.},
``The Complete Light-curve Sample of Spectroscopically Confirmed SNe Ia from Pan-STARRS1 and Cosmological Constraints from the Combined Pantheon Sample'',
Astrophys.\ J.\  {\bf 859} (2018) no.2,  101,
doi:10.3847/1538-4357/aab9bb
[arXiv:1710.00845 [astro-ph.CO]];
The numerical data of the full Pantheon SnIa sample are available at
http://dx.doi.org/10.17909/T95Q4X,
https://archive.stsci.edu/prepds/ps1cosmo/index.html.


\end{thebibliography}

\end{document}